\documentclass[12pt,oneside]{article}
\usepackage{amsfonts,amssymb,graphicx}
\def\be{\begin{equation}}
\def\ee{\end{equation}}
\def\bea{\begin{eqnarray}}
\def\eea{\end{eqnarray}}
\def\<{\langle}
\def\>{\rangle}
\def\~{\tilde}
\def\s{\sigma}

\def\a{\alpha}
\def\b{\beta}
\def\g{\gamma}

\def\t{\tau}

\newcommand{\brac}[1]{\< #1\>}
\newcommand{\av}[1]{\mbox{{\rm Av}}\left(#1\right)}
\newcommand{\dete}[1]{\mbox{det}\left(#1\right)}

\newtheorem{remark}{Remark}

\newtheorem{theorem}{THEOREM}
\newtheorem{definition}{Definition}

\begin{document}

\begin{center}{\sc\Large Thermodynamical Limit for Correlated Gaussian
Random Energy Models}
\end{center}
\begin{center}{ P. Contucci, M. Degli Esposti,
C. Giardin\`a, S. Graffi}\\
{\small Dipartimento di Matematica} \\
    {\small Universit\`a di Bologna,
    40127 Bologna, Italy}\\
    {\small {e-mail:
    $\{$contucci,desposti,giardina,graffi$\}$@dm.unibo.it}}
\end{center}
\hfill{\it To Francesco Guerra on his sixtieth birthday $\qquad$}
%
%
\begin{abstract}\noindent
Let  $\{E_{\s}(N)\}_{\s\in\Sigma_N}$ be a family of $|\Sigma_N|=2^N$
centered unit Gaussian  random variables defined by the covariance
matrix $C_N$ of elements $\displaystyle
 c_N(\s,\tau):=\av{E_{\s}(N)E_{\tau}(N)}$, and $H_N(\s) = -
\sqrt{N} E_{\s}(N)$ the  corresponding random Hamiltonian.  Then the
quenched thermodynamical limit exists if, for every decomposition
$N=N_1+N_2$, and all pairs $(\s,\t)\in \Sigma_N\times \Sigma_N$:
$$
c_N(\s,\tau)\leq \frac{N_1}{N}\;c_{N_1}(\pi_1(\s),\pi_1(\tau))+
\frac{N_2}{N}\;c_{N_2}(\pi_2(\s),\pi_2(\tau))
$$
where $\pi_k(\s), k=1,2$ are the projections of $\s\in\Sigma_N$
into $\Sigma_{N_k}$. The condition is explicitly verified for
the Sherrington-Kirckpatrick,  the even $p$-spin, the
Derrida REM and the Derrida-Gardner GREM models.
\end{abstract}
\section{Introduction, Definitions and Results}
It has recently been proved by Guerra and Toninelli \cite{GuTo}
that for the Sherrington-Kirckpatrick (hereafter SK) model (as well
as for the even-$p$-spin models) the thermodynamical limit exists
for the quenched free energy and almost everywhere for its random
realizations. In this paper we single out general sufficient
conditions that imply the existence of the quenched
thermodynamical limit for any correlated Gaussian random energy
model. Our analysis thus includes as special cases not only the even
$p$ spin models (in particular the SK
one, $p=2$) but also the Derrida REM model\cite{De1},\cite{De2} and the
Derrida-Gardner GREM\cite{DeGa}.

The paper is organized as follows: in this section we introduce
the definitions and state the results. In section 3, after
introducing and elucidating the operation of {\it lifting} for a
family of Gaussian random variables, we describe the proof of our
theorem. In section 4 we show how our analysis  applies to the
specific examples listed above.

To define the set up we consider a disordered model having $2^N$
energy levels where $N$ is the size of the system. We label the
energy levels by the index $\s = \{\s_1,\s_2,\ldots,\s_N\}$ where
each $\s_i$ takes the values $\pm 1$ for $i=1,\ldots,N$. We denote
$\Sigma_N$ the set of all $\s$. Then $|\Sigma_N|=2^N$. Clearly
$\Sigma_N$ coincides with the space of all  possible $2^N$
Ising configurations of length $N$.
\begin{definition}
\label{modello} Denote $\{E_{\s}(N)\}_{\s\in\Sigma_N}$  a family of
$2^N$ {\em centered unit Gaussian} random variables:
\be
\label{centered} 
\av{E_{\s}(N)} = 0 \; ,
\ee 
and covariance matrix
$C_N$ with elements defined by
\begin{eqnarray}
\label{unit} c_N(\s,\s) :=
\av{E^2_{\s}(N)} = 1 \; ,
\\
 c_N(\s,\t) := \av{E_{\s}(N)E_{\tau}(N)} \; .
\end{eqnarray}
 Here $\av{-}$
denotes expectation with respect to the probability measure
\be 
dP\left(E_1,\ldots,E_{2^{N}}\right)
\,=\,\frac{1}{\sqrt{(2\pi)^{2^N}\dete{C}}}\,\,\,
e^{-\frac{1}{2}\<E,\,C^{-1}E\>}\,\,dE_1\cdots dE_{2^{N}} .
\ee
\end{definition}

\begin{definition}
\par\noindent
\begin{enumerate}
\item
For each $N$ the Hamiltonian is given by
\be 
H_N(\s) =
-\sqrt{N}E_{\s}(N) \; .
\label{ee} 
\ee 
\item 
The partition
function of the system is:
\be 
\label{part}
Z_N(\b,E) = \sum_{\s} e^{-\b
H_N(\s)} = \sum_{\s} e^{\b\sqrt{N} E_{\s}(N)}
\ee 
\item
The  {\em quenched} free energy
$f_N(\b)$ of the system is defined as:
\be 
-\b f_N(\b) := \a_N(\b) := \frac{1}{N}\;
\av{\ln Z_N(\b,E)} \; .
\label{fe}
\ee
\end{enumerate}
\end{definition}
\begin{remark}
{\rm From now on we write $ E_{\s}(N)= E_{\s}$, dropping the
$N$-dependence. Remark moreover 
that Definition 1 includes Gaussian families of the form
\begin{eqnarray}
E_\s(N) \, = \, J_0 + \sum_{i}J_i\s_i + \sum_{i,j}J_{i,j}\s_i\s_j +
\sum_{i,j,k}J_{i,j,k}\s_i\s_j\s_k +
\nonumber
\\
+\ldots 
+\sum_{i_1,i_2,...,i_N}J_{i_1,i_2,...,i_N}\s_{i_1}\s_{i_2}...\s_{i_N} \;
\label{genercrem}
\end{eqnarray}
in which every $J$ is an indipendent Gaussian variable.
}
\end{remark}
{\bf Examples}.
\begin{enumerate}
\item The SK model.
Consider first the model defined by
\be 
\label{SKN}
 E_{\s}
:=\frac{1}{N}\sum_{i,j=1}^N J_{i,j}\s_i\s_j
\ee 
where the $J_{i,j}$ are $N^2$ i.i.d. unit Gaussian random variables.
A short computation yields
$$
{\rm Av}(E_{\s}E_{\t})=[q_N(\s,\t)]^2
$$
where, as usual 
\be 
\label{overlap}
q_N(\s,\t):=\frac{1}{N}\sum_{k=1}^N \s_k\t_k \label{chiu}
\ee 
is the
overlap between the $\s$ and $\t$ spin configurations. The standard SK model
is instead defined by
\be 
\label{SK}
 E_{\s}^{SK}
:=\frac{1}{N}\sum_{i<j=1}^N J_{i,j}\s_i\s_j \; .
\ee 
However the quenched free energy densities (\ref{fe}) of the two models
coincide up to a rescaling of the temperature, i.e.:
\be
\a_N^{SK}(\sqrt{2}\beta) \, = \, \a_N (\b) \, ,
\label{id}
\ee
In fact, 
 $J_{i,j}\s_i\s_j$ are centered, unit and i.i.d. Gaussian
random variables $\forall\,(i,j)$, and $J_{i,j}\s_i\s_j= J_{j,i}\s_j\s_i$.
Hence
$\displaystyle J_{i,j}\s_i\s_j+J_{j,i}\s_j\s_i\,
\mathrel{\mathop{=}\limits^{\cal D}}\,\sqrt{2}J_{i,j}\s_i\s_j$ (here
$\displaystyle
\mathrel{\mathop{=}\limits^{\cal D}}$ denotes equality in distribution of
two random variables).  Therefore, taking into account also the $N$
diagonal terms: 
\be 
\sqrt{N}E_{\s} \, \mathrel{\mathop{=}\limits^{\cal D}}\,
\sqrt{N}\sqrt{2}E^{SK}_{\s} + J \; ,
\label{lkj}
\ee 
where $J$ is a centered unit Gaussian variable. By
(\ref{part},\ref{fe}) formula (\ref{lkj}) immediately yields the relation
(\ref{id}).
\item 
The $p$-spin models. Here we consider the model:
\be 
\label{SKp}
E_{\s}:=\sqrt{\frac{1}{N^p}}\sum_{i_1,\ldots,i_p=1}^NJ_{i_1,\ldots,i_p}\s_{i
_1
}\cdots\s_{i_p} 
\ee 
where the $J_{i_1,\ldots,i_p}$ are once more i.i.d. unit
Gaussian random variables. As before, a short computation yields
\be 
{\rm Av(E_{\s}E_{\t})}=[q_N(\s,\t)]^p
\ee
\item 
The Derrida REM. Here the model is specified by Definition
\ref{modello} with 
\be 
{\rm Av}(E_{\s}E_{\t})=\delta(\s,\t)
\ee
\item The Derrida-Gardner GREM. Its inclusion into the above framework is
described in detail in Section 3.3.
\end{enumerate}
\begin{definition}
\label{def:pai}
For each $\s\in\Sigma_N$ let $\pi_1$ and $\pi_2$ be the two
canonical projections over the two  subsets $\Sigma_{N_1}$
and
$\Sigma_{N_2}$, generated by a partition ${\cal P}$ of the coordinates
$(\s_1,\ldots,\s_N)$ into a subset of $N_1$ coordinates and into a
complementary set of $N_2$ coordinates:
$N_1+N_2=N$, 
$\Sigma_N=\Sigma_{N_1}\times\Sigma_{N_2}$,
$\pi_1\otimes\pi_2=1_{\Sigma_N}$.
\end{definition}
\noindent
(Example:  $N=4$;  $\sigma\in \Sigma_4$ with coordinates denoted
$\{\sigma_1, \s_2, \s_3, \s_4 \}$. Consider for $N_1=N_2=2$ the partition
 ${\cal P}\s=(\s_1,\s_2)\cup (\s_3,\s_4)$. Then
$\Sigma_N=\Sigma_{N_1}\times\Sigma_{N_2}$ and the two projections
$\pi_k:\Sigma_N\to \Sigma_{N_k}$, $k=1,2$ act in the following
way: 
$\pi_1(\sigma_1, \s_2,
\s_3,
\s_4) = (\sigma_1, \s_2)$ and $\pi_2(\sigma_1, \s_2, \s_3, \s_4) =
(\sigma_3, \s_4)$).
\par\noindent 
Our main result is the following:
\begin{theorem}
\label{th:gnocca}
Let the covariance
matrices $C_N$ fulfill the condition: \be c_N(\s,\t) \,-
\,\frac{N_1}{N}\;c_{N_1}(\pi_{1}(\s),\pi_{1}(\t)) \, - \,
                 \frac{N_2}{N}\;c_{N_2}(\pi_{2}(\s),\pi_{2}(\t)) \; \le
\; 0
\; , \label{punct} \ee
for every $N\ge \tilde{N}$, every $(\s,\t)\in \Sigma_N\times
\Sigma_N$ and every decomposition $N_1+N_2=N$. Then the thermodynamical
limit exists, 
in the sense that
\be
\lim_{N\to\infty}\frac{1}{N}{\rm Av}(\log Z_N(\beta))
=\sup_{N}\frac{1}{N}{\rm Av}(\log Z_N(\beta)) \; . \label{gnocca}
\ee
\end{theorem}
\begin{remark}
{\rm The  result $(\ref{gnocca})$ can be extended to the
almost-everywhere convergence of free energy density, internal
energy and ground state energy with elementary probability methods
(see $\cite{GuTo}$)}
\end{remark}
\begin{remark}
{\rm The conditions (\ref{punct}) are not }necessary. {\rm The proof
itself will show that we only need the sign of the quantity in the
left hand side of (\ref{punct}) in average, not pointwise.
Moreover the condition (\ref{centered}) can be replaced by a more
general small deviation vanishing for large $N$ and (\ref{unit}) by a
uniform (in N) bound over the diagonal terms. We plan to return
over such a general case elsewhere.}
\end{remark}
\begin{remark}
{\rm It is still an open interesting question whether the class of models
 we control the thermodynamical limit of do have, in that
limit, the properties axiomatically introduced by Ruelle in
$\cite{rugrem}$ to define directly the infinite particle systems.
To this purpose see $\cite{BS}$, $\cite{BK1}$, $\cite{BK2}$ and
$\cite{BK3}$.}
\end{remark}

\section{Proof}
Within this section it is useful to consider $2$ identical {\it
copies} of the same system: the system $1$ is assigned the
Hamiltonian $H(\s)$ and the system $2$ the Hamiltonian $H(\t)$.

\vspace{0.2cm} 
\noindent
\begin{definition}
The {\it quenched} measure over the two copies $\<-\>$ is defined
by 
\be <-> \, = \, {\rm Av}{[Z(\beta,E)]^{-2}\sum_{(\s,\t)\in
\Sigma_N\times \Sigma_N} - \; e^{\beta (H(\s)+H(\t))}} \, .
\label{bracket} 
\ee 
The definition may of course be generalized to
$r$ copies.
\end{definition}

We want now to embed a Gaussian system $\{E_\s\}_{\Sigma_K}$ into
a larger one $\{E_\t\}_{\Sigma_L}$ for some $K<L$. In particular
we want to embed two of them of size $N_1$ and $N_2$ into one of
size $N=N_1+N_2$. Our embedding procedure is defined in terms of
the two canonical projections $\pi_j$, $j=1,2$ from $\Sigma_N$ to
$\Sigma_{N_j}$ given in Definition (\ref{def:pai}).
\begin{definition}
Given the family $\{E_\mu\}_{\Sigma_{N_1}}$ of size $N_1$ we {\rm
lift} it to one of size $N$: $\{E^{(1)}_\s\}_{\Sigma_{N}}$
defining 
\be E^{(1)}_{\s} \, \mathrel{\mathop{=}\limits^{\cal D}}
\, E_{\pi_1(\s)} \; .
\ee 
Moreover starting from
$\{E_\mu\}_{\Sigma_{N_2}}$ we define in the same way
$\{E^{(2)}_\s\}_{\Sigma_{N}}$ by \be E^{(2)}_{\s} \,
\mathrel{\mathop{=}\limits^{\cal D}} \, E_{\pi_2(\s)} \; . \ee
Having defined each family $\{E_\s\}_{\Sigma_{N}}$,
$\{E^{(1)}_\s\}_{\Sigma_{N_1}}$ and
$\{E^{(2)}_\s\}_{\Sigma_{N_2}}$ we specify their joint
distribution requiring mutual independence.
\end{definition}
\begin{remark}: 
{\rm The embedded Gaussian
systems $\{E^{(1)}_\s\}_{\Sigma_{N_1}}$ and
$\{E^{(2)}_\s\}_{\Sigma_{N_2}}$ are degenerate: In fact for all
$\s$ and $\t$ such that $\pi_1(\s) = \pi_1(\t)$ \be E^{(1)}_\s \,
= \, E^{(1)}_\t \; . \ee Summarizing we define the joint measure
of $\{E_\s\}_{\Sigma_{N}}$, $\{E^{(1)}_\s\}_{\Sigma_{N_1}}$ and
$\{E^{(2)}_\s\}_{\Sigma_{N_2}}$ $d\hat{P}=dP dP_{1} dP_{2}$
defined by the three covariances $C_{N}$, $C_{N_1}$ and $C_{N_2}$.}
\end{remark}
{\bf Proof of THEOREM 1.} \\
We proceed in three lemmas. \\
\noindent
Lemma 0. \hskip .3truecm {\bf Interpolation}
\par\noindent
Given a pair $(\pi_1,\pi_2)$ as before,  following \cite{GuTo}, we
 pick three {\it independent} Gaussian systems
$E^{(j)}_{\pi_j(\sigma)}$, $j=0,1,2$ and introduce the quantity
($\pi_0(\sigma)=\sigma$)

\be
\label{zt} 
H_{(N,N_1,N_2)}(\sigma,t) \; := \;
-\sum_{j=0}^2\sqrt{t_jN_j} E_{\pi_j(\sigma)}^{(j)}
\ee 
where
$t_0=t$ and $t_1=t_2=(1-t)$, and the correspondent partition sum
\be 
\label{zt2} 
Z_N(t,\beta) \;: = \; \sum_{\sigma\in\Sigma_N}
e^{-\beta H_{(N,N_1,N_2)}(\sigma,t)}.
\ee
It is now easy to see that:
\be
 Z_N(1,\beta)=Z_N{(\beta)} \; ,
\label{int1} 
\ee 
and
\begin{eqnarray}
Z_N(0,\beta)&=&\sum_{\sigma\in\Sigma_N} e^{\beta (
\sqrt{N_1}E_{\pi_1(\sigma)}^{(1)} \,+
\,\sqrt{N_2}E_{\pi_2(\sigma)}^{(2)})}\nonumber\\
&=&\sum_{\tau\in\Sigma_{N_2}}\,\,\sum_{\sigma\in\Sigma_N;
\,\,\pi_2(\sigma)=\tau} e^{\beta (\sqrt{N_1}E_{\pi_1(\sigma)}^{(1)} \,+
\,\sqrt{N_2}E_{\tau}^{(2)})}\nonumber\\
&=&\sum_{\tau\in\Sigma_{N_2}}\,e^{\beta
\sqrt{N_2}E_{\tau}^{(2)}}\,\sum_{\g\in\Sigma_{N_1}}
e^{\beta \sqrt{N_1}E_{\g}^{(1)}}\nonumber\\
&=&Z_{N_1}(\beta)\cdot Z_{N_2}(\beta)\label{interpolo}
\end{eqnarray}
\noindent
Lemma 1. \hskip .3truecm {\bf Boundedness}\\
The Jensen inequality
\be 
\av{\log Z} \, \le \, \log(\av{Z})
\ee
implies 
\be 
\frac{1}{N}\av{\log Z_N(\beta)}
\; \le \; \log(2)+\frac{\beta^2}{2}
\ee
because by (\ref{part}) $\displaystyle \av{Z}=2e^{\b^2/2}$ after
performing the Gaussian integration.
 \noindent 
Lemma 2. \hskip
.3truecm {\bf Monotonicity}\\Taking the $t$ derivative of the
logarithm of (\ref{zt2}) we get: (here we abbreviate $H_{N,N_1,N_2}=H$)

\be \frac{d}{dt}\log
Z_N(t)\,=\,\frac{\beta}{Z_N(t)}\,\sum_{\sigma\in\Sigma_N}\left(\sum_{k=0}^2
\epsilon_k \sqrt{\frac{N_k}{t_k}} E_{\pi_k(\sigma)}^{(k)}
\,e^{-\beta H(\sigma,t)}\right)\, , \ee where $\epsilon_0=1$ and
$\epsilon_1=\epsilon_2=-1$.

We now use the integration by parts formula for correlated Gaussian
variables $\{\xi_i\}$ with covariance $c_{i,j}$, which states
\be
\av{\xi_j\cdot f}\,=\,\av{\sum_{k=1}^n c_{j,k}\cdot\frac{\partial
f}{\partial \xi_k}} \; .
\ee 
This yields
\begin{eqnarray}
\label{sum} \av{\frac{1}{\beta }\frac{d}{dt}\log Z_N(t)}=
\sum_{\sigma\in\Sigma_N}\sum_{k=0}^2\epsilon_k\,\sqrt{\frac{N_k}{t_k}}\,
\av{\frac{E_{\pi_k(\sigma)}^{(k)}\,e^{-\beta H}}{Z_N(t)}}\\
=\sum_{\sigma\in\Sigma_N}\sum_{k=0}^2\epsilon_k\,\sqrt{\frac{N_k}{t_k}}\,
\av{ \sum_{\tau_k\in\Sigma_{N_k}}
c_{N_k}(\pi_k(\sigma),\tau_k)\cdot\frac{\partial}{\partial
E_{\tau_k}^{(k)}}\frac{e^{-\beta H}}{Z_N(t)}}\nonumber
\end{eqnarray}

\noindent Given now $\tau_k \in\Sigma_{N_k}$ fixed, we calculate

\begin{eqnarray}
&&\frac{\partial}{\partial E_{\tau_k}^{(k)}}\frac{e^{-\beta
H(\sigma,t)}}{Z_N(t)}  =
\beta\frac{\sqrt{N_kt_k}\;\delta_{\tau_k}^{\pi_k(\sigma)}\;e^{-\beta
H(\sigma,t)}\cdot Z_N(t)-e^{-\beta H(\sigma,t)}\cdot\frac{\partial
Z_N}{\partial
E_{\tau_k}^{(k)}}}{Z_N^2(t)} \nonumber \\
\,\nonumber\\ & = &
\beta\frac{\sqrt{N_kt_k}\;\delta_{\tau_k}^{\pi_k(\sigma)}
e^{-\beta H(\sigma,t)}\cdot Z_N(t)-\sqrt{N_kt_k}\;e^{-\beta
H(\sigma,t)}\cdot\sum_{\xi\in\Sigma_N,\,\pi_k(\xi)=\tau_k}
e^{-\beta H(\xi,t)}}{Z_N^2(t)}\nonumber
\end{eqnarray}

\noindent 
The term with $k=0$ in formula (\ref{sum}) is easy to
calculate and we get:
\begin{eqnarray}
&&
N\beta\av{\sum_{\sigma\in\Sigma_N}\sum_{\tau\in\Sigma_N}c_N(\sigma,\tau)
\left[\delta_\tau^\sigma \frac{e^{-\beta H(\sigma,t)}}{Z_N}-
\sum_{\xi\in\Sigma_N} \delta_\xi^\tau e^{-\beta(H(\xi,t)+
H(\sigma,t))}\right]}\,=\,\nonumber\\
 &=& N\beta \av{\sum_{\sigma\in\Sigma_N}
c_N(\sigma,\sigma)\cdot\frac{e^{-\beta H(\sigma,t)}}{Z_N}\;-\;
 \sum_{(\s,\t)\in \Sigma_N\times \Sigma_N}
 c_N(\sigma,\tau)\,e^{-\beta(H(\tau,t)+H(\sigma,t))}}\nonumber\\
 &=& N\beta \brac{1\,-\, c_N(\s,\t)}_t \; ,
\end{eqnarray}
where $<->_t$ is the quenched measure with respect to the
Hamiltonian (\ref{zt}).
\vspace{0.3cm} \noindent
In the same way
for the term $k=1$ (and similarly for $k=2$) we obtain:
\begin{eqnarray}
N_1\beta\av{\sum_{\sigma\in\Sigma_N}
\sum_{\tau\in\Sigma_{N_1}}c_{N_1}(\pi_1(
\sigma),\tau)\left[ \delta^{\tau}_{\pi_1(\sigma)}\frac{e^{-\beta
H(\sigma,t)}}{Z_N}- \sum_{\xi\in\Sigma_N} \delta_{\pi_1(\xi)}^\tau
e^{-\beta(H(\xi,t)+ H(\sigma,t))}\right]}\,=\, \nonumber
\\
= N_1\brac{1\,-\, c_{N_1}(\pi_{1}(\s),\pi_1(\t))}_t
\; . \qquad\qquad\qquad\qquad\qquad\qquad\qquad
\label{sbbc}
\end{eqnarray}
Summing up the three contributions we obtain:
\begin{eqnarray}
\frac{1}{N}\frac{d}{dt}\av{\log Z_N(t)} = \qquad\qquad\qquad
\qquad\qquad\nonumber
\\
 = -{\beta^2}<c_N(\s,\t) -
\frac{N_1}{N}c_{N_1}(\pi_{1}(\s),\pi_{1}(\t)) -
                 \frac{N_2}{N}c_{N_2}(\pi_{2}(\s),\pi_{2}(\t)) >_t \; ,
\label{der}
\end{eqnarray} 
and, by the hypothesis (\ref{punct}):
\be 
\frac{d}{dt}\av{\log
Z_N(t)} \ge 0 \; . \label{pos}
\ee 
Formula (\ref{pos}) together with
the boundary conditions (\ref{int1}) and (\ref{interpolo}) gives
for every $N_1+N_2=N$
\be 
\a_N \ge \frac{N_1}{N}\a_{N_1} +
\frac{N_2}{N}\a_{N_2} \; .
\label{sub}
\ee 
This entails Theorem \ref{th:gnocca} as
explained for instance in \cite{Ru2}.
\begin{remark}
{\rm  Lemma 3 is indeed a particular
case
of a theorem by J-P. Kahane \cite{K} (see also \cite{LT}, Theorem
3.11, p.74). The Gaussian process $X$ of \cite{K} can in fact be
identified with our Gaussian process $\sqrt{N}E$,  and the process
$Y$ with our process $\sqrt{N_1}E^{(1)}+\sqrt{N_2}E^{(2)}$. The
further identifications  $A\equiv\Sigma_N\times\Sigma_N$,
$B=\emptyset$, $f\equiv
\ln{Z}$ 
immediately entail that  Hypothesis (1) of \cite{K} reduces to
(\ref{punct}) and Assertion (3) to our formula (\ref{sub}) because
Hypothesis (2) is just convexity of $\ln{Z}$.}
\end{remark} 
\section{Examples} \subsection{The SK and even $p$-spin
models} For the sake of completeness we recover here the
Guerra-Toninelli result \cite{GuTo}.
First note that
by the definition $(\ref{chiu})$ we have
\be 
q_N(\s,\t) \,- \,\frac{N_1}{N}q_{N_1}(\pi_{1}(\s),\pi_{1}(\t))
\, - \,
                 \frac{N_2}{N}q_{N_2}(\pi_{2}(\s),\pi_{2}(\t))
\; = \; 0 \;
. \label{rfm} 
\ee 
so that $(\ref{punct})$ holds as an
equality for $p=1$ (the random field model). By $(\ref{sub})$ this
means that the random field model free energy density doesn't depend
on the size: $\a_N = \a_1$. For $p=2u$ (SK corresponds to $u=1$)
formula $(\ref{rfm})$ together with the convexity of the
function $x\to x^{2u}$ implies  $(\ref{punct})$:
\be
q^{2u}_N(\s,\t) \,-
\,\frac{N_1}{N}q^{2u}_{N_1}(\pi_{1}(\s),\pi_{1}(\t)) \, - \,
                 \frac{N_2}{N}q^{2u}_{N_2}(\pi_{2}(\s),\pi_{2}(\t)) \; \le
\; 0 \; . \label{evenp}
\ee
For the standard $p$-spin model defined as
\be 
\label{SKp1}
 E_{\s}=\sqrt{\frac{p!}{2N^p}}\sum_{i_1<\ldots <
i_p}J_{i_1,\ldots,i_p}\s_{i_1
}\cdots\s_{i_p} 
\ee 
we refer to \cite{GuTo}
\subsection{The REM}
The model is defined by:
\be \av{E_{\s}E_{{\s}'}} =
\delta_{\s,\s'} . 
\ee  
Condition $(\ref{punct})$ is verified
because it becomes 
\be 
\delta_{\s,\s'} \leq
\frac{N_1}{N}\delta_{\pi_1(\s),\pi_1(\s')} + \frac{N_2}{N}
\delta_{\pi_2(\s),\pi_2(s')} \; .
\ee 
In fact if $\s=\s'$ the
previous formula is an identity. If $\s\neq\s'$ the left hand side
is $0$ but the right hand side is not always zero.  Let us take for
instance $\s = (+,+)$ and $\s'=(+,-)$, $\pi_1(+,+)=+$,
$\pi_1(+-)=+$, $\pi_2(+,+)=+$, $\pi_2(+,-)=-$. In that case the
left hand side is zero and the right hand side is $1/2$.

\subsection{The GREM}
To show the inclusion in our scheme of the Derrida-Gardner GREM
\cite{DeGa} let us first  recall its construction. The GREM considers
$2^N$ Gaussian random energies $H(\mu) = \sqrt{N}E_\mu$. Their covariance
is specified after the assignment of a {\it rooted tree} with $n$ layers
and
$2^N$ leaves, $n<N$. The root {\it furcates} into 
$\alpha_1^N$ branches, the
vertices at the end of the first layer {\it furcate}
into $\alpha_2^N$ branches  etc.,  up to the vertices at the end
of the $n-1$ layer which $\alpha_n^N$-{\it furcate} into the $2^N$
leaves. 
\begin{remark}
{\rm The topological constraint over the successive furcations which
end up on $2^N$ leaves implies $\prod_{i=1}^{n}\alpha_{i}^{N}=2^N$.
Each $\alpha_i^N$ is an integer which by the previous formula
divides $2^N$. By the fundamental theorem of arithmetics  
$\alpha_{i}^{N}=2^{k_i}$. Here $k_i, i=1,\ldots,n$ is a non-negative
integer, and
$k_1 + k_2 + ... k_n =N$. In other words: given  any tree with
$2^N$ leaves the  construction  allows only for furcations in powers of 2 
at each layer. The cofficients $\alpha_i$
must depend on $N$: in fact, $\alpha_{i}=2^{\frac{k_i}{N}}$ and the only
$N$-independent
choice of the vector $\alpha$ is obtained for $k_i=\frac{N}{l_i}$ where
the integers
$l_i$ have to divide $N$ {\it for all $N$}. Hence they must fulfill the
constraint
$\sum_{i=1}^{n}\frac{1}{l_i}=1$ which is impossible.}
\end{remark}
The previous remark  allows
us to associate to each leave $\mu$ a
spin configuration $\{\s_1, \s_2, ..., \s_N\}$. This can be
done observing that the $\alpha_1(N)^N=2^{k_1(N)}$ branches emerging
from the root identify canonically the configurations of $k_1$
spins, the successive branches the configuration of $k_2$ spins
and so on. We have in this way associated to each leaf either a
path (the only one joining the root to it) or a spin
configuration. The model is finally specified by the formula
$E(\mu) = \sum_{i=1}^{n} \epsilon_{i}^{(\mu)}$ where the
$\epsilon_i$ are thrown according to $n$ Gaussians with 
${\rm Av}(\epsilon_i)=0$ and ${\rm Av}[(\epsilon_i)^2]=a_i$: to each
branch of the tree we associate an independent $\epsilon$ whose
distribution depends (through its variance)  only at which layer
starts the branch. Defining $\displaystyle
v^{(l)}=\sum_{i=1}^{l-1}a_i\,,\;\;$ ($v^{(0)}=0$ and
$v^{(1)}=1$) it is immediate to prove that if two paths $\mu$ and
$\nu$ merge at the level $l$ we have ${\rm Av}(E_\mu E_\mu) = v^{(l)}$.
For fixed $n$ and $N$ this construction is exactly the
Derrida-Gardner process over a tree ${\cal T}_{n,N}$; we will denote
it $\{{\cal E}, {\cal T}_{n,N}\}$.

Theorem 1 entails  existence of the thermodynamical limit for the GREM, in
the sense that if  $\{{\cal E}, {\cal T}_{n,N}\}$ is assigned for a given
$n$ and all $N>n$, and the sequence of $\{{\cal T}_{n,N}\}$ is {\it
increasing} i.e.
$k_i(N)\ge k_i(M)$ for $N\ge M$ then
 its free energy density is (at fixed n) decreasing (and bounded) in
$N$. To show this assertion, starting from a process $\{{\cal E}, {\cal
T}_{n,N_1}\}$ we build the process $\{{\cal E}^{(1)}_{\pi_1}, {\cal
T}_{n,N}\}$ with
$N=N_1+N_2$ in the following way: at each vertex of the tree
${\cal T}_{n,N_1}$ sitting on the layer $i$ we increase the
multiplicity of the {\it furcation} by a factor $2^{k_i(N)-k_i(N_1)}$
assigning {\rm the same value} $\epsilon_i^{(1)}$ to all newly
introduced branches. By construction the new process will enjoy the
property 
\be
{\rm Av}(E^{(1)}_{\pi_1(\sigma)}E^{(1)}_{\pi_1(\tau)}) \ge v^{(l)} \; .
\label{c1} 
\ee 
We apply the same construction to build $\{{\cal
E}^{(2)}_{\pi_2}, {\cal T}_{n,N}\}$ and we have
\be
{\rm Av}(E^{(2)}_{\pi_2(\sigma)}E^{(2)}_{\pi_2(\tau)}) \ge v^{(l)} \; .
\label{c2} 
\ee 
It is now straightforward to verify that
conditions $(\ref{c1})$ and $(\ref{c2})$ imply $(\ref{punct})$.
\vskip 12pt\noindent
{\small {\bf Acknowledgments.}  
One of us (P.C.) thanks
Francesco Guerra for useful conversations and Michael Aizenman for
introducing him to the Correlated Gaussian Random Energy Models.
We also thank the referees and Anton Bovier for interesting observations
and for pointing out the reference \cite{K}.
\par\noindent
This work has been partially supported by the EC
RTN-HPRN-CT-2000-00103  (Mathematical Aspects of
Quantum Chaos) and by 
Universit\`a di Bologna, Funds for Selected Research Topics.}

\end{document}